\documentclass[prl,aps,twocolumn,superscriptaddress,groupaddress,showpacs,preprintnumbers]
{revtex4-1}

\usepackage{dcolumn}
\usepackage{bm}
\usepackage{graphics}
\usepackage{graphicx}
\usepackage{placeins}
\usepackage[colorlinks,breaklinks]{hyperref}

\newcommand{\gam}{\gamma}
\def\pom{{\rm P\kern -0.53em\llap I\,}}

\begin{document}
\preprint{FERMILAB-PUB-11-643-PPD}

\lefthyphenmin=2
\righthyphenmin=3

\title{
Observation of Exclusive {\boldmath $\gamma\gamma$} Production in
{\boldmath $p\bar{p}$} Collisions at {\boldmath $\sqrt{s}=1.96$}~TeV
}
\affiliation{Institute of Physics, Academia Sinica, Taipei, Taiwan 11529, Republic of China}
\affiliation{Argonne National Laboratory, Argonne, Illinois 60439, USA}
\affiliation{University of Athens, 157 71 Athens, Greece}
\affiliation{Institut de Fisica d'Altes Energies, ICREA, Universitat Autonoma de Barcelona, E-08193,
Bellaterra (Barcelona), Spain}
\affiliation{Baylor University, Waco, Texas 76798, USA}
\affiliation{Istituto Nazionale di Fisica Nucleare Bologna, $^{ee}$University of Bologna, I-40127
Bologna, Italy}
\affiliation{University of California, Davis, Davis, California 95616, USA}
\affiliation{University of California, Los Angeles, Los Angeles, California 90024, USA}
\affiliation{Instituto de Fisica de Cantabria, CSIC-University of Cantabria, 39005 Santander, Spain}
\affiliation{Carnegie Mellon University, Pittsburgh, Pennsylvania 15213, USA}
\affiliation{Enrico Fermi Institute, University of Chicago, Chicago, Illinois 60637, USA}
\affiliation{Comenius University, 842 48 Bratislava, Slovakia; Institute of Experimental Physics,
040 01 Kosice, Slovakia}
\affiliation{Joint Institute for Nuclear Research, RU-141980 Dubna, Russia}
\affiliation{Duke University, Durham, North Carolina 27708, USA}
\affiliation{Fermi National Accelerator Laboratory, Batavia, Illinois 60510, USA}
\affiliation{University of Florida, Gainesville, Florida 32611, USA}
\affiliation{Laboratori Nazionali di Frascati, Istituto Nazionale di Fisica Nucleare, I-00044
Frascati, Italy}
\affiliation{University of Geneva, CH-1211 Geneva 4, Switzerland}
\affiliation{Glasgow University, Glasgow G12 8QQ, United Kingdom}
\affiliation{Harvard University, Cambridge, Massachusetts 02138, USA}
\affiliation{Division of High Energy Physics, Department of Physics, University of Helsinki and
Helsinki Institute of Physics, FIN-00014, Helsinki, Finland}
\affiliation{University of Illinois, Urbana, Illinois 61801, USA}
\affiliation{The Johns Hopkins University, Baltimore, Maryland 21218, USA}
\affiliation{Institut f\"{u}r Experimentelle Kernphysik, Karlsruhe Institute of Technology, D-76131
Karlsruhe, Germany}
\affiliation{Center for High Energy Physics: Kyungpook National University, Daegu 702-701, Korea;
Seoul National University, Seoul 151-742, Korea; Sungkyunkwan University, Suwon 440-746, Korea;
Korea Institute of Science and Technology Information, Daejeon 305-806, Korea; Chonnam National
University, Gwangju 500-757, Korea; Chonbuk National University, Jeonju 561-756, Korea}
\affiliation{Ernest Orlando Lawrence Berkeley National Laboratory, Berkeley, California 94720, USA}
\affiliation{University of Liverpool, Liverpool L69 7ZE, United Kingdom}
\affiliation{University College London, London WC1E 6BT, United Kingdom}
\affiliation{Centro de Investigaciones Energeticas Medioambientales y Tecnologicas, E-28040 Madrid,
Spain}
\affiliation{Massachusetts Institute of Technology, Cambridge, Massachusetts 02139, USA}
\affiliation{Institute of Particle Physics: McGill University, Montr\'{e}al, Qu\'{e}bec, Canada
H3A~2T8; Simon Fraser University, Burnaby, British Columbia, Canada V5A~1S6; University of Toronto,
Toronto, Ontario, Canada M5S~1A7; and TRIUMF, Vancouver, British Columbia, Canada V6T~2A3}
\affiliation{University of Michigan, Ann Arbor, Michigan 48109, USA}
\affiliation{Michigan State University, East Lansing, Michigan 48824, USA}
\affiliation{Institution for Theoretical and Experimental Physics, ITEP, Moscow 117259, Russia}
\affiliation{University of New Mexico, Albuquerque, New Mexico 87131, USA}
\affiliation{The Ohio State University, Columbus, Ohio 43210, USA}
\affiliation{Okayama University, Okayama 700-8530, Japan}
\affiliation{Osaka City University, Osaka 588, Japan}
\affiliation{University of Oxford, Oxford OX1 3RH, United Kingdom}
\affiliation{Istituto Nazionale di Fisica Nucleare, Sezione di Padova-Trento, $^{ff}$University of
Padova, I-35131 Padova, Italy}
\affiliation{University of Pennsylvania, Philadelphia, Pennsylvania 19104, USA}
\affiliation{Istituto Nazionale di Fisica Nucleare Pisa, $^{gg}$University of Pisa,
$^{hh}$University of Siena and $^{ii}$Scuola Normale Superiore, I-56127 Pisa, Italy}
\affiliation{University of Pittsburgh, Pittsburgh, Pennsylvania 15260, USA}
\affiliation{Purdue University, West Lafayette, Indiana 47907, USA}
\affiliation{University of Rochester, Rochester, New York 14627, USA}
\affiliation{The Rockefeller University, New York, New York 10065, USA}
\affiliation{Istituto Nazionale di Fisica Nucleare, Sezione di Roma 1, $^{jj}$Sapienza
Universit\`{a} di Roma, I-00185 Roma, Italy}
\affiliation{Rutgers University, Piscataway, New Jersey 08855, USA}
\affiliation{Texas A\&M University, College Station, Texas 77843, USA}
\affiliation{Istituto Nazionale di Fisica Nucleare Trieste/Udine, I-34100 Trieste, $^{kk}$University
of Udine, I-33100 Udine, Italy}
\affiliation{University of Tsukuba, Tsukuba, Ibaraki 305, Japan}
\affiliation{Tufts University, Medford, Massachusetts 02155, USA}
\affiliation{University of Virginia, Charlottesville, Virginia 22906, USA}
\affiliation{Waseda University, Tokyo 169, Japan}
\affiliation{Wayne State University, Detroit, Michigan 48201, USA}
\affiliation{University of Wisconsin, Madison, Wisconsin 53706, USA}
\affiliation{Yale University, New Haven, Connecticut 06520, USA}

\author{T.~Aaltonen}
\affiliation{Division of High Energy Physics, Department of Physics, University of Helsinki and
Helsinki Institute of Physics, FIN-00014, Helsinki, Finland}
\author{M.G.~Albrow}
\affiliation{Fermi National Accelerator Laboratory, Batavia, Illinois 60510, USA}
\author{B.~\'{A}lvarez~Gonz\'{a}lez$^z$}
\affiliation{Instituto de Fisica de Cantabria, CSIC-University of Cantabria, 39005 Santander, Spain}
\author{S.~Amerio}
\affiliation{Istituto Nazionale di Fisica Nucleare, Sezione di Padova-Trento, $^{ff}$University of
Padova, I-35131 Padova, Italy}
\author{D.~Amidei}
\affiliation{University of Michigan, Ann Arbor, Michigan 48109, USA}
\author{A.~Anastassov$^x$}
\affiliation{Fermi National Accelerator Laboratory, Batavia, Illinois 60510, USA}
\author{A.~Annovi}
\affiliation{Laboratori Nazionali di Frascati, Istituto Nazionale di Fisica Nucleare, I-00044
Frascati, Italy}
\author{J.~Antos}
\affiliation{Comenius University, 842 48 Bratislava, Slovakia; Institute of Experimental Physics,
040 01 Kosice, Slovakia}
\author{G.~Apollinari}
\affiliation{Fermi National Accelerator Laboratory, Batavia, Illinois 60510, USA}
\author{J.A.~Appel}
\affiliation{Fermi National Accelerator Laboratory, Batavia, Illinois 60510, USA}
% ...
\collaboration{CDF Collaboration}\thanks{With visitors from
$^a$Istituto Nazionale di Fisica Nucleare, Sezione di Cagliari, 09042 Monserrato (Cagliari), Italy,
$^b$University of CA Irvine, Irvine, CA 92697, USA,
$^c$University of CA Santa Barbara, Santa Barbara, CA 93106, USA,
$^d$University of CA Santa Cruz, Santa Cruz, CA 95064, USA,
$^e$Institute of Physics, Academy of Sciences of the Czech Republic, Czech Republic,
$^f$CERN, CH-1211 Geneva, Switzerland,
$^g$Cornell University, Ithaca, NY 14853, USA,
$^h$University of Cyprus, Nicosia CY-1678, Cyprus,
$^i$Office of Science, U.S. Department of Energy, Washington, DC 20585, USA,
$^j$University College Dublin, Dublin 4, Ireland,
$^k$ETH, 8092 Zurich, Switzerland,
$^l$University of Fukui, Fukui City, Fukui Prefecture, Japan 910-0017,
$^m$Universidad Iberoamericana, Mexico D.F., Mexico,
$^n$University of Iowa, Iowa City, IA 52242, USA,
$^o$Kinki University, Higashi-Osaka City, Japan 577-8502,
$^p$Kansas State University, Manhattan, KS 66506, USA,
$^q$Korea University, Seoul, 136-713, Korea,
$^r$University of Manchester, Manchester M13 9PL, United Kingdom,
$^s$Queen Mary, University of London, London, E1 4NS, United Kingdom,
$^t$University of Melbourne, Victoria 3010, Australia,
$^u$Muons, Inc., Batavia, IL 60510, USA,
$^v$Nagasaki Institute of Applied Science, Nagasaki, Japan,
$^w$National Research Nuclear University, Moscow, Russia,
$^x$Northwestern University, Evanston, IL 60208, USA,
$^y$University of Notre Dame, Notre Dame, IN 46556, USA,
$^z$Universidad de Oviedo, E-33007 Oviedo, Spain,
$^{aa}$CNRS-IN2P3, Paris, F-75205 France,
$^{bb}$Texas Tech University, Lubbock, TX 79609, USA,
$^{cc}$Universidad Tecnica Federico Santa Maria, 110v Valparaiso, Chile,
$^{dd}$Yarmouk University, Irbid 211-63, Jordan.
}
\noaffiliation

\date{September 21, 2011}

\begin{abstract}
We have observed exclusive $\gamma\gamma$ production in proton-antiproton
collisions at $\sqrt{s}=1.96$~TeV, using data from 1.11~$\pm$~0.07~fb$^{-1}$
integrated
luminosity taken by the Run~II Collider Detector at Fermilab. We selected events
with two electromagnetic showers, each with transverse energy $E_T > 2.5$ GeV
and pseudorapidity $|\eta| <$ 1.0, with no other particles detected in $-7.4 <
\eta < +7.4$. The two showers have similar $E_T$ and azimuthal angle
separation $\Delta\phi \sim \pi$; 34 events have two charged particle tracks,
consistent with the QED process $p \bar{p} \rightarrow p + e^+e^- + \bar{p}$ by
two-photon exchange, while 43 events have no charged tracks. The
number of these events that are exclusive $\pi^0\pi^0$ is consistent with zero
and is $<$ 15 at 95\% C.L. The cross section for $p\bar{p} \to
p+\gamma\gamma+\bar{p}$ with $|\eta(\gamma)| < 1.0$ and $E_T(\gamma) > 2.5$~GeV
is $2.48\,^{+0.40}_{-0.35}(\mathrm{stat})\,^{+0.40}_{-0.51}(\mathrm{syst})\,
\mathrm {
pb}$.
\end{abstract} 


\maketitle

\begin{figure}[htbp]
\begin{center}
\includegraphics[width=0.38\textwidth]{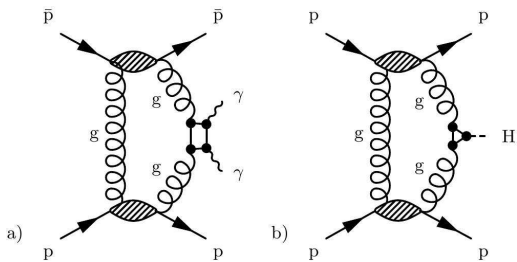}
\caption{Leading order diagrams for central exclusive production
in $p(\bar{p})-p$ collisions: a) exclusive $\gamma\gamma$ production 
in $\bar{p}-p$ collisions; b) exclusive Higgs boson production in $p-p$
collisions.
Note the screening gluon that cancels the color flow from the
interacting gluons.
\label{fig:cepDiagrams} }
\end{center}
\end{figure}

In proton-(anti)proton collisions, two direct high-$E_T$ photons can be produced
at leading order by $q\bar{q}$$\,\rightarrow\,$$\gam\gam$ and by 
$gg$$\,\rightarrow\,$$\gam\gam$
through a quark loop. In the latter case it is possible for another gluon
exchange to cancel the color of the fusing gluons, allowing the (anti)proton to
emerge intact with no hadrons produced. For $p\bar{p}$ collisions, this is the
``exclusive" process
$p\bar{p}$$\,\rightarrow\,$$p$$\,+\,$$\gamma\gamma$$\,+\,$$\bar{p}$, 
for which the leading order
diagram is shown in Fig.~\ref{fig:cepDiagrams}a~\cite{cdfloi,exclgg1}. The outgoing
(anti)proton has nearly the
beam momentum, and transverse momentum $p_T \lesssim 1$ GeV/c, 
having emitted a pair of
gluons in a color singlet. There is a pseudorapidity gap
$\Delta\eta >$ 6
adjacent to the (anti)proton. In Regge theory this is diffractive scattering
via pomeron~\cite{forshawross,donnachie}, $\rm I\kern +0.53em\llap P$, exchange.
The cross section for $|\eta(\gamma)| < 1.0$ and transverse energy $E_T(\gamma)
> 2.5$~GeV is predicted~\cite{KMRgg,KMRgg2} to be
$\sigma(\gam\gam)_{\mathrm{exclusive}} \sim$~0.2 - 2 pb, depending on the
low-$x$ (unintegrated) gluon density. 
Additional uncertainties come from the cross section
for $g+g$$\,\rightarrow\,$$\gam+\gam$, the probability that no hadrons are
produced by additional
parton interactions (rapidity gap survival factor and Sudakov
suppression~\cite{sudakov}), and
the probability that neither proton dissociates (e.g., $p$$\,\rightarrow\,$$p\
\pi^+\pi^-$)~\cite{KMRgg}.
The calculation is also imprecise because of the low $Q^2$, the squared 4-momentum transfer.
The total theoretical uncertainty on the cross section can 
be estimated to be a factor $^{\times{3}}_{\div{3}}$ \cite{exchiggs}.
Apart from its
intrinsic interest for QCD, the process tests the theory of exclusive Higgs
boson production~\cite{cdfloi,exclgg1,KMRgg,bialaslandshoff,schafer,
exchiggs,harland,forshaw,acf}
$p+p$$\,\rightarrow\,$$p+H+p$, Fig.~\ref{fig:cepDiagrams}b, which may be
detectable at the LHC.
The leading order processes $gg$$\,\rightarrow\,$$\gamma\gamma$ and
$gg$$\,\rightarrow\,$$H$
are calculable perturbatively, but the more uncertain elements of the exclusive
processes (mainly the unintegrated gluon densities, the Sudakov suppression, and
the gap survival probability) are common to both (see
Fig.~\ref{fig:cepDiagrams}). For a 120 GeV standard model Higgs boson the
exclusive cross section at $\sqrt{s} = 7$~TeV is 3~fb with a factor
$^{\times{3}}_{\div{3}}$ uncertainty \cite{exchiggs}.

Processes other than $gg$$\,\rightarrow\,$$\gamma\gamma$ can produce an
exclusive
$\gamma\gamma$ final state. Contributions from
$q\bar{q}$$\,\rightarrow\,$$\gamma\gamma$ and 
$\gamma\gamma$$\,\rightarrow\,$$\gamma\gamma$ are respectively $<$
5\% and $<$ 1\% of $gg$$\,\rightarrow\,$$\gamma\gamma$ \cite{KMRgg}. Backgrounds
to
exclusive $\gamma\gamma$ events to be considered are $\pi^0\pi^0$ and
$\eta\eta$, with each meson decaying to two photons, of which one is not
detected. We also consider events where one or both protons dissociate, e.g., 
$p$$\,\rightarrow\,$$p\,\pi^+\pi^-$, to be background. These
backgrounds are small.

We previously published a search for exclusive $\gamma\gamma$
production, finding three candidate events with $E_T(\gam) >$ 5 GeV and $|\eta|
<$ 1.0, using data from 532~pb$^{-1}$ of integrated luminosity \cite{cdfgg1}.
The prediction of Ref.~\cite{KMRgg} was 0.8$^{+1.6}_{-0.5}$ events.
Two events had a single narrow electromagnetic (EM) shower on each side, as
expected for $\gam\gam$, but no observation could be claimed. This Letter
reports the observation of 43 events with a contamination of $< 15 \:
\pi^0\pi^0$ events (at 95\% C.L.), after we lowered the trigger threshold on the
EM showers from 4 GeV to 2 GeV and collected data from another
1.11~fb$^{-1}$ of integrated luminosity. We used the QED process
$p$$\,+\,$$\bar{p}$$\,
$$\rightarrow$$\,$$p\,$$+$$\,\gamma^*\gamma^*$$+$$\,\bar{p}$$\,
$$\rightarrow$$\,$$p$$\,+\,$$e^+e^-$$+\,$$\bar{p}$ in the same
data
set, for which the cross section is well known, as a check of the analysis.

The data were collected by the Collider Detector at Fermilab, CDF II, at the
Tevatron, with
$p\bar{p}$ collisions at $\sqrt{s}$ = 1.96 TeV. The CDF II detector is a general
purpose detector described elsewhere~\cite{cdf}; here we give a brief summary of
the detector components used in this analysis. Surrounding the beam pipe is a
tracking system consisting of a silicon microstrip detector, a cylindrical drift
chamber (COT)~\cite{cot}, and a solenoid providing a 1.4~Tesla magnetic field. 
The tracking system is fully efficient at reconstructing isolated tracks with
$p_T \geq$ 1 GeV/c and $|\eta|<1$.  It is surrounded by the central and end-plug
calorimeters covering the range $|\eta|<3.6$.  Both calorimeters have separate
EM and hadronic compartments. A proportional wire chamber
(CES)~\cite{balka}, with orthogonal anode wires and cathode strips, is embedded
in the central EM calorimeter, covering the region of $|\eta| < 1.1$, at a depth
of six radiation
lengths. It allows a measurement of the number and shape, in both $\eta$ and
azimuth $\phi$, of EM showers (clusters of wires or strips).
The anode-wire pitch (in $\phi$) is 1.5~cm
and the cathode-strip pitch varies with $\eta$ from 1.7~cm to 2.0~cm. The CES
provides a means of distinguishing single photon showers from $\pi^0 \rightarrow
\gamma\gamma$ up to $E_T(\pi^0) \sim$ 8 GeV. The region $3.6<|\eta|<5.2$ is
covered by a lead-liquid scintillator calorimeter called the
Miniplug~\cite{miniplug}. At higher pseudorapidities, $5.4<|\eta|<7.4$,
scintillation counters, called beam shower counters (BSC-1/2/3), are located on
each
side of the CDF detector. Gas Cherenkov detectors, with 48 photomultipliers per
side, covering 3.7 $<|\eta|<$ 4.7, detect charged particles, and were also used
to determine the luminosity with a 6\% uncertainty~\cite{clc}.

The data were recorded using a three-level on-line event selection system
(trigger). At the first level we required one EM cluster with $E_T >$
2 GeV and $|\eta| < 2.1$ and no signal above noise in the BSC-1 counters
($|\eta| = 5.4 - 5.9)$. This rapidity gap requirement rejected a large fraction
of inelastic collisions as well as most events with more than one interaction (pileup). 
A second EM cluster with similar properties was required at level two. A level
three trigger selected events with two calorimeter showers consistent with
coming from electrons or photons: i.e., passing the requirement (cut) that the
ratio of shower energy in the hadronic (HAD) calorimeter to that in the EM
(HAD:EM) be less than 0.125, and that the signal shape in the CES is
consistent with a single shower.  

We now describe the offline selection of events, with two isolated EM showers
and no other particles except the outgoing $p$ and $\bar{p}$, which were not
detected. Two central, $|\eta|<1$, EM showers were required with $E_T >$ 2.5 GeV to
avoid trigger threshold inefficiencies. The energy resolution is $dE/E \sim 8\%$ from 
test beam studies and \textit{in situ} $p/E$ matching for electrons. A refined HAD:EM
ratio cut of $< 0.055+0.000\:45E$ was applied, as well as an acoplanarity cut of
$|\pi-\Delta\phi|<0.6$.
The trigger selection efficiency for
single photons was measured using data collected with an interaction trigger
(minimum bias). The BSC-1 gap trigger was taken to be 100\% efficient as
the BSC-1 trigger threshold was clearly above the noise level and the
offline selection criteria. 
We measured an overall trigger efficiency of
$\varepsilon_{\mathrm{trig}}=92\%\pm2\%$(syst).
A weighting process was necessary due to the different slope in $E_T$ of the
minimum bias probe data compared to the signal.
The trigger efficiency did not show any $\eta$ or $\phi$ dependence for 
$|\eta|<1$. Monte Carlo signal simulation data samples were generated using the
\textsc{superchic} program (version 1.3)~\cite{harland,superchic} based on recent
developments of
the Durham KMR model~\cite{exclgg1}. The Monte Carlo samples were passed through
a simulation of the detector, \textsc{cdfsim}~6.1.4.m including \textsc{geant} 
version 3.21/14~\cite{geant}.
The systematic error was estimated by using the binwise uncertainty of the
efficiency in the weighting
process of the signal Monte Carlo sample.
Taking into account a combined detector and offline
reconstruction efficiency of $\varepsilon_{\mathrm{rec}}=55\%\pm3\%$(syst), and
a photon
identification efficiency of
$\varepsilon_{\mathrm{id}}=93\%\pm1\%$(syst), we obtained a photon-pair
efficiency 
$\varepsilon_{\mathrm{pho}}=\varepsilon_{\mathrm{trig}}^2*\varepsilon_{\mathrm{
rec } }*\varepsilon_{\mathrm{id}}^2=40\%\pm3\%$(syst).
The systematic
uncertainties of the reconstruction and
identification efficiency were estimated by shifting kinematical input
parameters over a reasonable interval motivated by the dominating 
EM-energy-scale uncertainty~\cite{escale}.
The offline selection then required that no activity other than these two
showers (or clusters of showers) occurred in the entire detector, $|\eta| < 7.4$.
We used the same procedure as in our earlier study of exclusive $e^+e^-$
events~\cite{cdfee}, searching all the calorimeters for any signal above noise
levels, determined using noninteraction events; 99.2\% of such events have no tower (out of 480)
with $E_T > 125$~MeV. We also required the CLC
counters and the more forward BSC counters to have signals consistent with only
noise. Events triggered only on a bunch crossing (zero bias) showed
that the exclusive efficiency, $\varepsilon_{excl}$, defined as the factor
to be applied to the delivered luminosity to account for the requirement of no
pileup, is
$\varepsilon_{excl} = 6.8\%\pm0.4\%$(syst). The probability $P(0)$ of a zero-bias
event satisfying all the exclusivity cuts, i.e., having no detected inelastic
interaction, is $P(0)=A\exp(-\bar{n})=A\exp(-L_{\mathrm{x}} \sigma_{\mathrm{vis}})$,where
$L_{\mathrm{x}}$ is the single bunch crossing luminosity (cm$^{-2}$) and $\sigma_{\mathrm{vis}}$ is
the visible cross section; $\sigma_{\mathrm{vis}}=\sigma_{\mathrm{inel}}$ if everyinelastic
collision is detected. We find $\sigma_{\mathrm{vis}}=67 \pm 6$~mb. In the absence of noise (above
our chosen thresholds) $A = 1.0$; we find $A = 0.98 \pm 0.02$.
We checked that
the rate of candidate events, corrected for the exclusive efficiency, was
constant during data taking (one year). The systematic uncertainty was
estimated using the spread in slope parameters from fits to data
in different time periods.

\begin{table}[!ht]\centering
\caption{ Summary of
parameters used for the measurement of the exclusive
photon-pair cross section for $E_{T}(\gamma)>$ 2.5 GeV and $|\eta(\gamma)| <
1.0$.
Values for the $e^+e^-$ control study
are also given. Note that b/g stands for
background.\label{summarystatgg}}\vspace{0.2cm}
\begin{tabular}{lc}\hline\hline
Integrated luminosity $\mathcal{L}_\mathsf{int}$& $1.11\pm0.07$
fb$^{-1}$\\
Exclusive efficiency & $0.068\pm0.004\,\mathrm{(syst)}$\\\hline \hline
Exclusive $\gamma\gamma$&  \\ \hline 
Events & 43 \\
Photon-pair efficiency& $0.40\pm0.02\,\mathrm{(stat)}\pm0.03\,\mathrm{(syst)}$\\
Probability of no conversions &$0.57\pm0.06$ (syst)\\
$\pi^0\pi^0$ b/g (events) & 0.0, $< 15$ (95\% C.L.)\\
Dissociation b/g (events)  & $0.14\pm0.14\,\mathrm{(syst)}$\\\hline \hline
Exclusive $e^+e^-$&  \\ \hline
Events & 34 \\
Electron-pair efficiency&
$0.33\pm0.01\,\mathrm{(stat)}\pm0.02\,\mathrm{(syst)}$\\
Probability of no radiation &$0.42\pm0.08\,\mathrm{(syst)}$\\
Dissociation b/g (events)  & $3.8\pm0.4\,\mathrm{(stat)}\pm0.9\,\mathrm{
(syst)}$\\\hline\hline
\end{tabular}
\end{table}

\begin{figure}[htbp]
\begin{center}
\begin{tabular}{cc}
\includegraphics[width=0.24\textwidth]{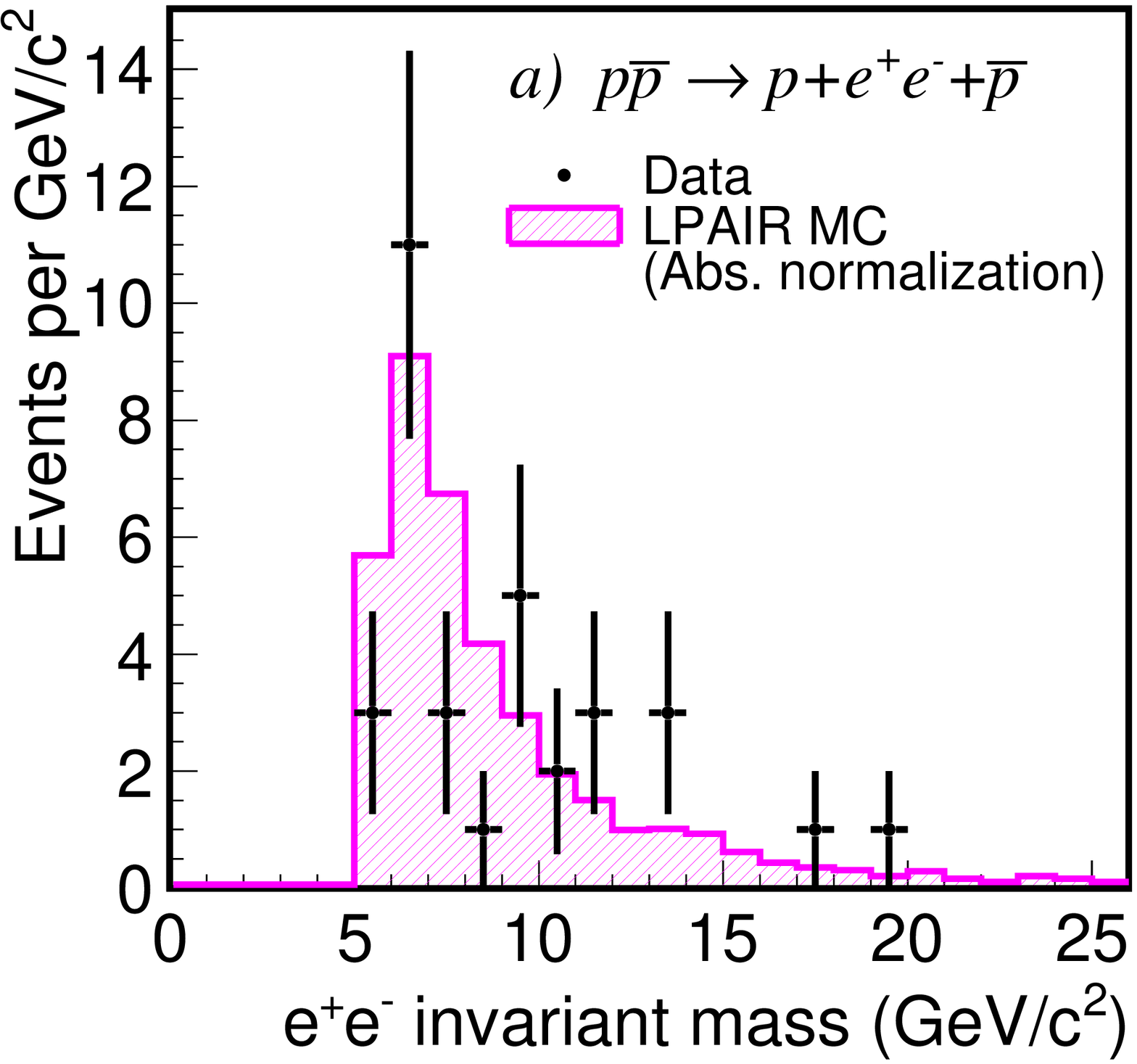}
&
\includegraphics[width=0.24\textwidth]{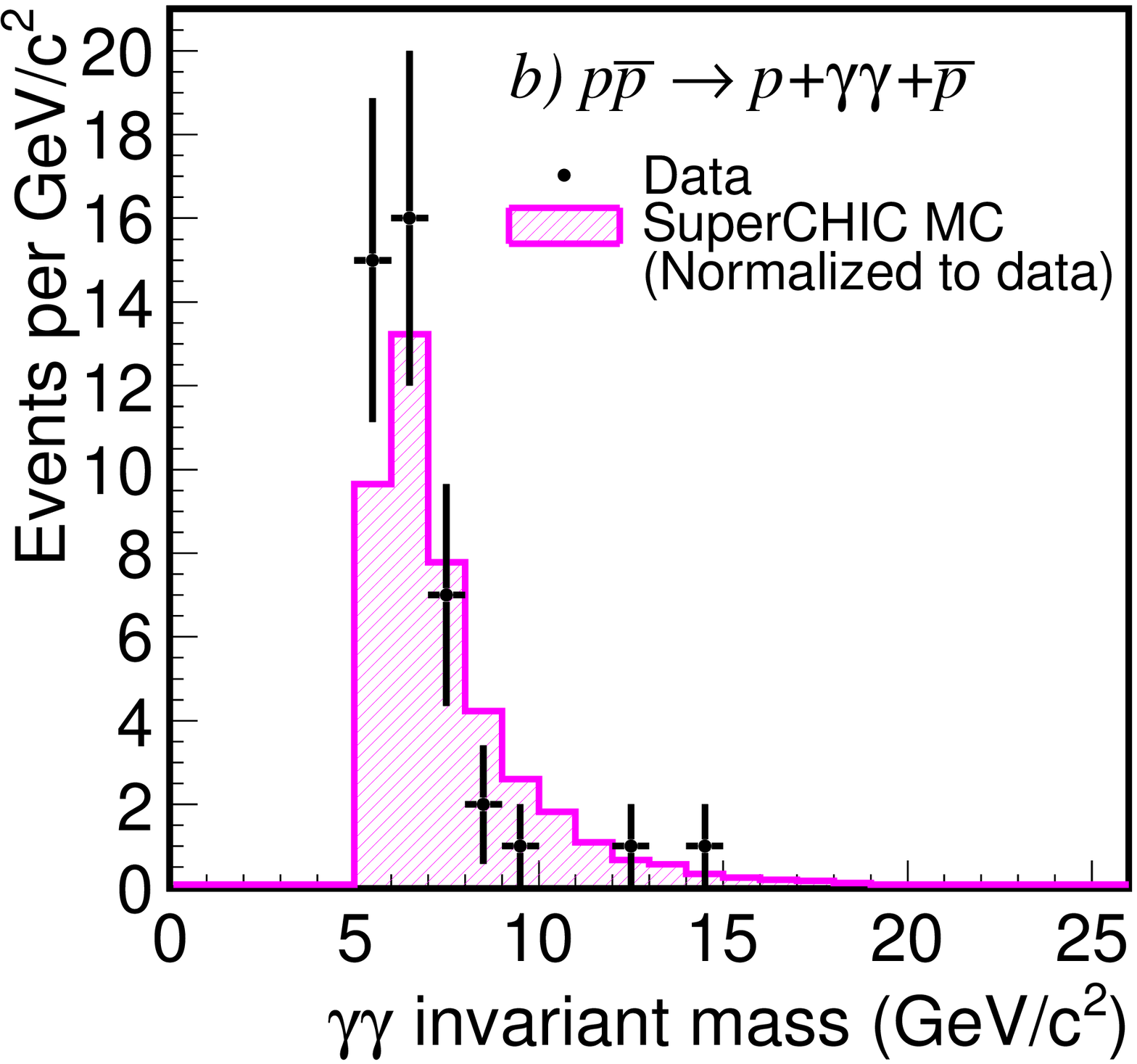}
\\
\includegraphics[width=0.24\textwidth]{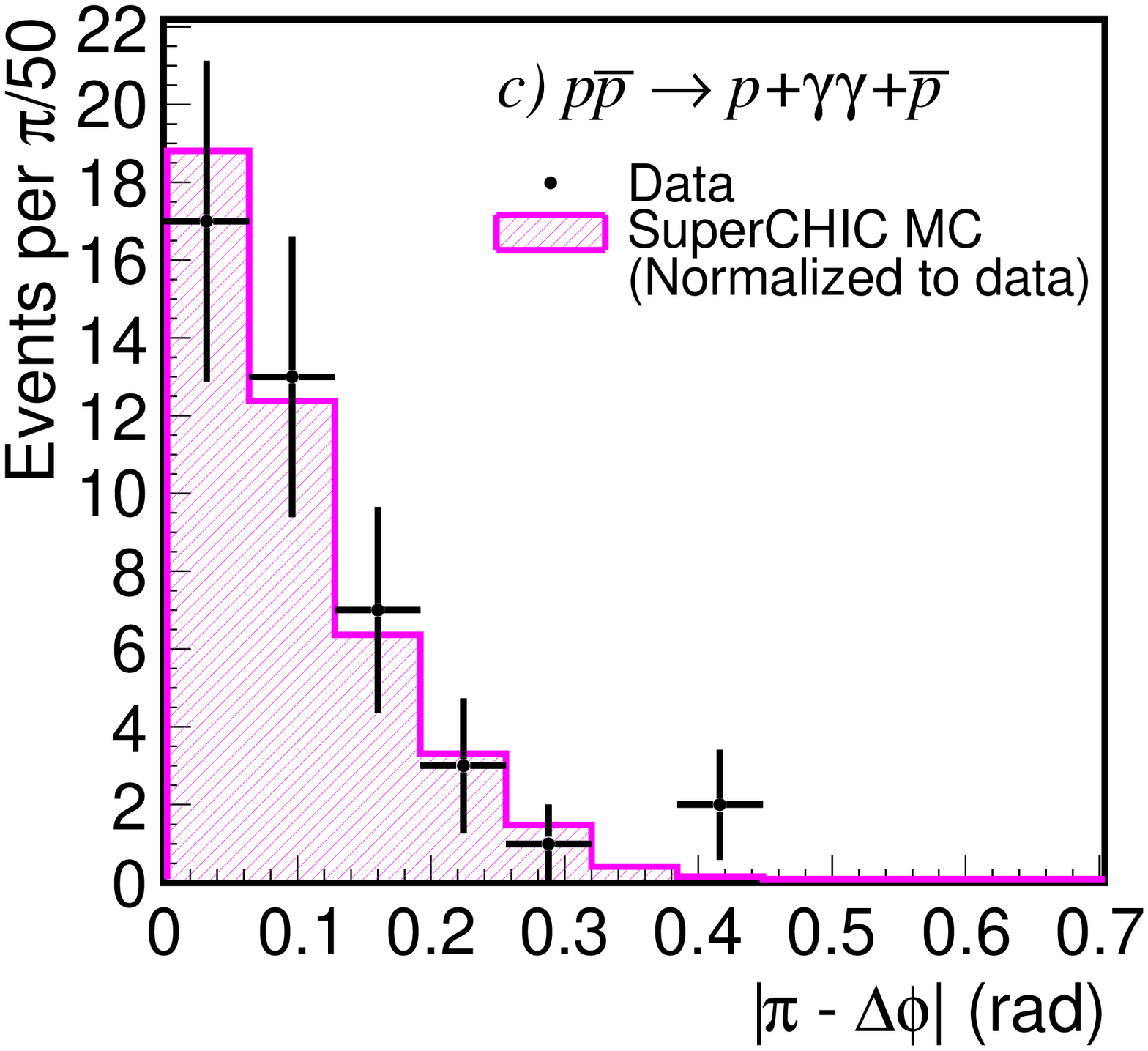}
&
\includegraphics[width=0.24\textwidth]{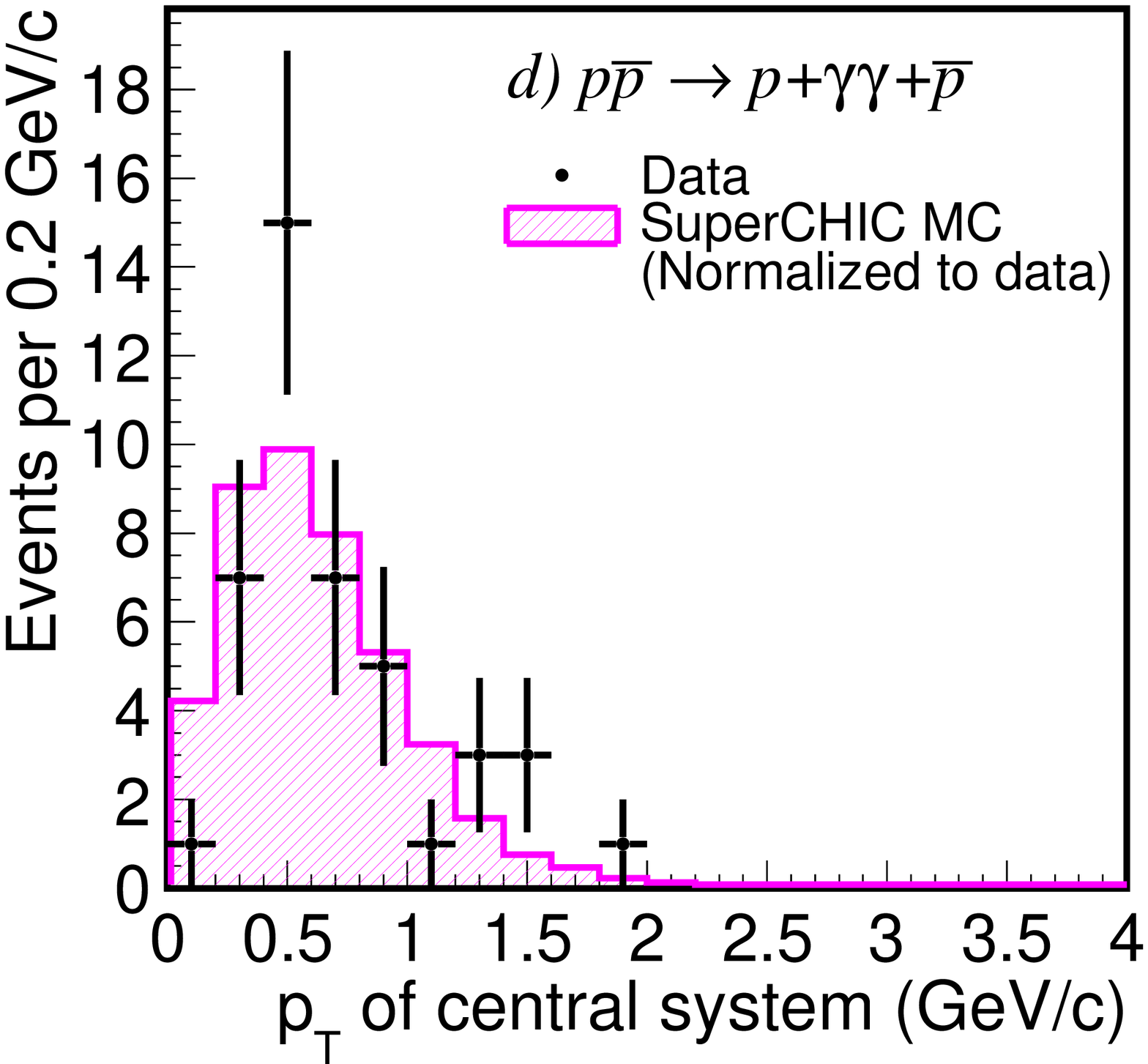}
\\
\end{tabular}
\caption{The $e^+e^-$ candidates: invariant mass distribution (a).
The two-photon candidates: invariant mass distribution (b), $|\pi-\Delta\phi|$
distribution (c), and $p_T$ distribution of the two photons (d). 
All error bars are statistical. The MC predictions for $\gamma\gamma$ are 
normalized to data. The QED prediction for $e^+e^-$ is normalized to the 
delivered luminosity and efficiencies. The MC samples for the QED process were 
generated with the \textsc{lpair} program~\cite{lpair}.
\label{fig:kinGGEE} }
\end{center}
\end{figure}

The selection of 81 events passing all cuts was made without reference to the
track detectors. We found that 34 have exactly two oppositely charged tracks, 43
have no tracks in the COT, and four are in neither class. Visual inspection of
the latter showed that two had photon conversions, and two were likely to be
$e^+e^-$ events with bremsstrahlung. These numbers are consistent with
expectations from the detector simulation.
The tracks in the 34 two-track events agree in all aspects with the
QED process $p+\bar{p} \rightarrow p + e^+e^- + \bar{p}$ via two virtual
photons, previously observed in CDF~\cite{cdfee,cdfz}. 
The calorimeter shower energies are consistent with the momenta measured from
the tracks.
Kinematic distributions, after detector simulation, are as expected.
The mass $M(e^+e^-)$ distribution is presented in Fig.~2a, together 
with the QED prediction normalized to the delivered luminosity and efficiencies, 
showing that the cross section agrees with the QED prediction in both 
magnitude and shape. We measured a 
cross section of  
$\sigma_{e^+e^-,\mathrm{exclusive}}(|\eta(e)|<1,\,E_T(e)>2.5\,\mathrm{GeV}) =
2.88\,^{+0.57}_{-0.48}(\mathrm{stat})\pm0.63(\mathrm{syst})$~pb,
compared to $3.25\pm0.07$~pb (QED,~\cite{lpair}). The systematic uncertainties
for the QED study are mostly identical to the photon case. 
Distinct from photons, electrons leave tracks in the tracking detectors and may
radiate.
The systematic uncertainty on the radiation probability was estimated by varying
the exclusivity cuts by $\pm10$\%.
This $e^+e^-$ sample provides a valuable check of the exclusive $\gamma\gamma$
analysis.

\begin{figure}[htbp]
\begin{center}
\begin{tabular}{cc}
\includegraphics[width=0.24\textwidth]{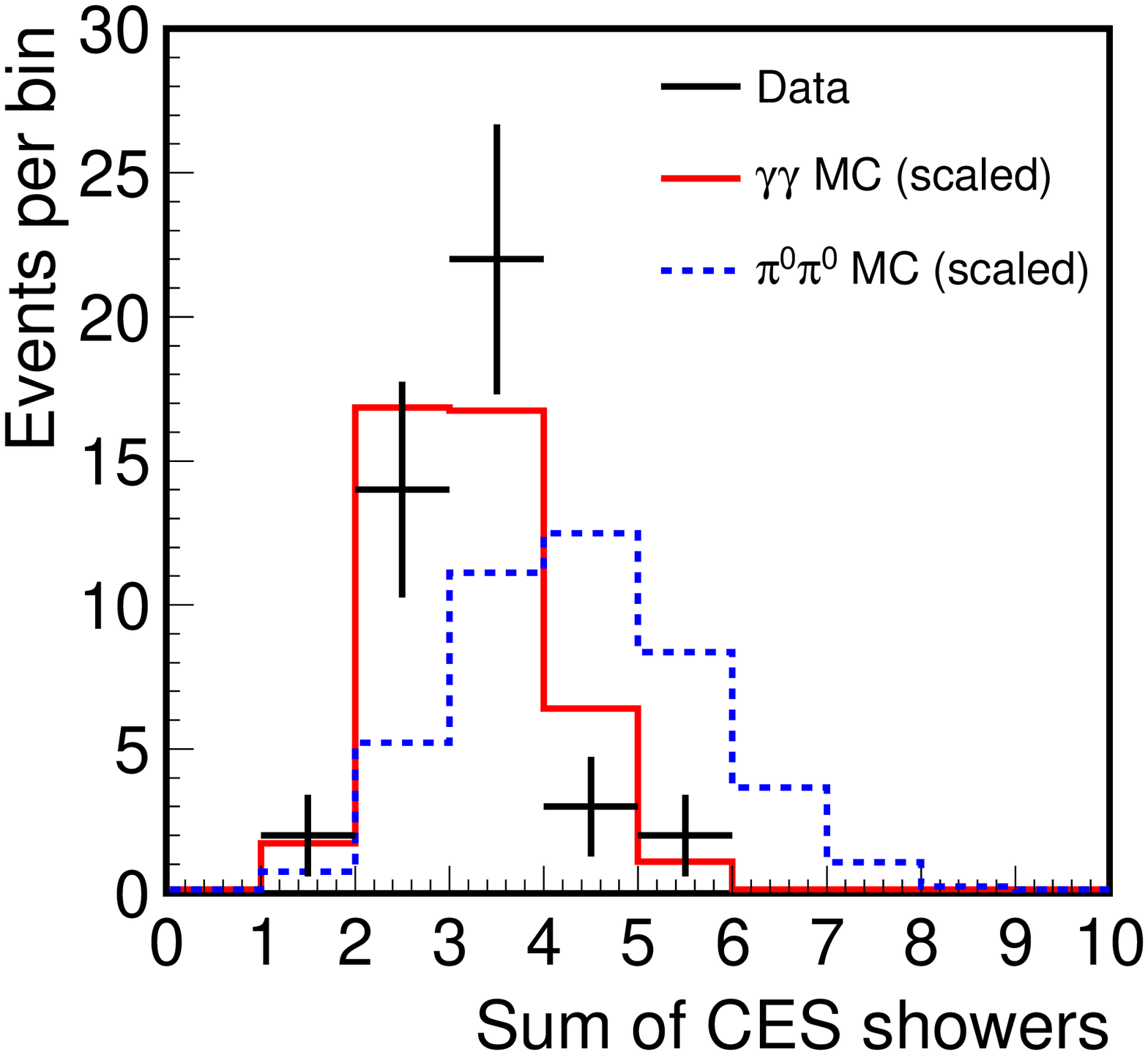}
&
\includegraphics[width=0.24\textwidth]{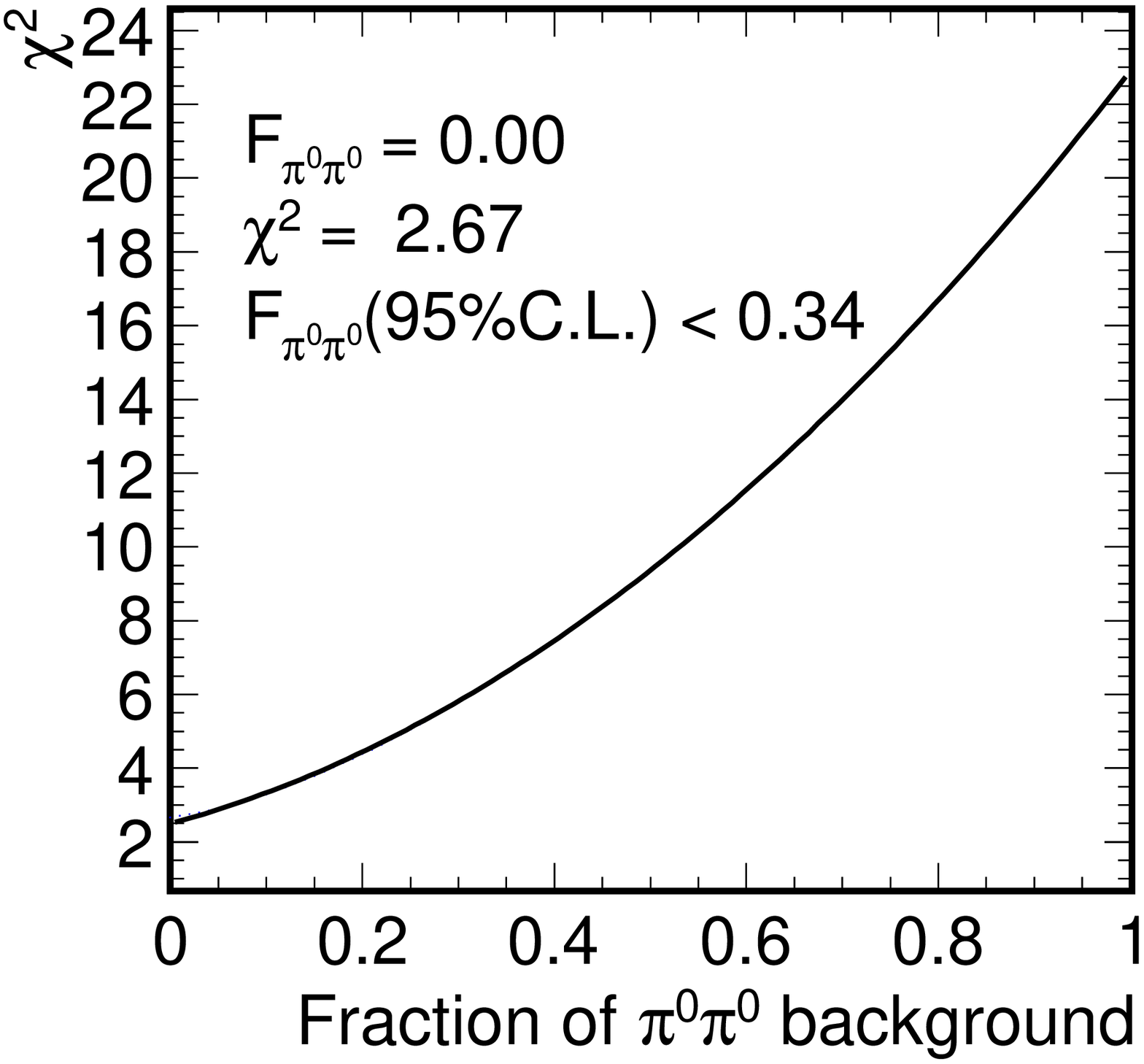}
\\
a)&b)\\
\end{tabular}
\caption{Estimate of $\pi^0\pi^0$ background fraction in the candidate
sample. Distribution of reconstructed CES showers per event for data
compared to $\gamma\gamma$ and $\pi^0\pi^0$ Monte Carlo simulations (a). Background
fraction estimate using Pearson's $\chi^2$ test to fit the composition
hypothesis to the data distribution (b).
\label{fig:pi0BG} }
\end{center}
\end{figure}

\begin{figure}[htbp]
\begin{center}
\includegraphics[width=0.36\textwidth]{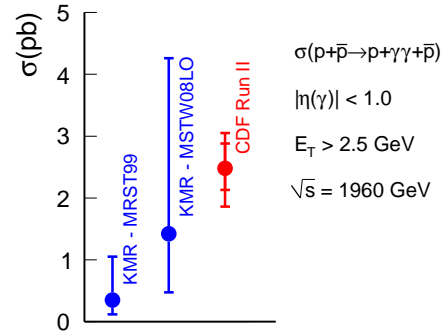}
\caption{Comparison of the measured cross section for the exclusive
$\gamma\gamma$
production in $p\bar{p}$ collisions at $\sqrt{s}=1.96$~TeV with theoretical 
predictions~\cite{harland}.\label{fig:expvsth}}       
 \end{center}
\end{figure}

The 43 events with no tracks have the kinematic properties 
expected for exclusive $\gamma\gamma$ production~\cite{superchic}. In particular
the $M(\gamma\gamma)$ distribution [Fig.~2b] extending up to 15 GeV/c$^2$ is as
expected, as well as the acoplanarity $\pi - \Delta \phi(\gamma\gamma)$
[Fig.~2c] and the 2-vector sum of $p_T$ [Fig.~2d];
in these plots [unlike Fig.~2a] the \textsc{superchic} Monte Carlo prediction is 
normalized to the same number of events as the data. An important issue is
whether some of these events could be $\pi^0\pi^0$, rather than $\gamma\gamma$.
Note that $\gamma\pi^0$ events are forbidden by \textit{C} parity. The CES chambers give
information on the number of EM showers. The minimum opening angle
$\Delta\theta_{min}$ between the two photons from $\pi^0$ decay is
$2\tan^{-1}\left(\frac{m(\pi)}{p(\pi)}\right)$ = 3.1$^\circ$ for $p(\pi)$ = 5
GeV, well separated in the CES chambers, which have a granularity $< 0.5^\circ$.
A $\pi^0$ can fake a $\gamma$ only if one photon ranges out before the CES, or
falls in an inactive region (8\%) of the detector. All of the 68 $e^{\pm}$
events in our sample, with similar energies, had matching showers in the CES
chambers.  A \textsc{geant}~\cite{geant} simulation predicts the probability
that a photon in our energy range produces a shower to be $\gtrsim98.3$\%.
We summed the number of reconstructed CES showers in the event, mostly 2 or 3 as
shown in Fig.~\ref{fig:pi0BG} (left). 
The distribution agrees very well with the $\gamma\gamma$ simulation, and
strongly disagrees with the $\pi^0\pi^0$ simulation. Fitting to the sum of the
two components gives a best fit to the fraction $F(\pi^0\pi^0)$ = 0.0, with a
95\% C.L. upper limit of 15 events. Since obtaining this result, a new
calculation of exclusive $\pi^0\pi^0$ production~\cite{hkms} predicts
$\sigma_{excl}(\pi^0\pi^0)$ = 6 - 24 fb for $E_T(\pi^0) > 2.5$ GeV and $|\eta|
< 1.0$, $\lesssim0.01$ of our measured exclusive $\gamma\gamma$ cross section.
In the cross section
calculation we take this background to be zero. Exclusive $\eta\eta$
production is also expected to be negligible.
The only other significant background could be undetected proton dissociation,
about 10\% for the QED $e^+e^-$ process but $<$1\% for~\,$\pom \:
+\,\pom\,\rightarrow \gam + \gam$~\cite{KMRgg,KMRchi,vkpriv}. 
The cross section for both photons with
$E_T(\gamma) > 2.5$~GeV and $|\eta(\gamma)| < 1.0$ and no other produced 
particles is given by:
\[
\sigma_{\gamma\gamma,\mathrm{exclusive}}
= \frac{N(candidates)
- N(background)}{\mathcal{L}_\mathsf{int} . \varepsilon .
\varepsilon_{excl}}\mathrm{,} 
\]
where $\varepsilon$ is the product of the trigger, reconstruction,
identification, and conversion efficiencies (22.8\%) in
Table~\ref{summarystatgg}. The systematic
uncertainty on the conversion probability was estimated by varying the
exclusivity cuts by $\pm10$\%. We find
$\sigma_{\gamma\gamma,\mathrm{excl}}\,(|\eta(\gamma)|<1,E_T(\gamma)>2.5~
\mathrm{GeV}) =
2.48\,^{+0.40}_{-0.35}(\mathrm{stat})\,^{+0.40}_{-0.51}(\mathrm{syst})$~pb. The
theoretical prediction~\cite{harland} is strongly dependent on the low-$x$
gluon
density, having central values 1.42 pb (\textsc{mstw08lo}) or 0.35 pb
(\textsc{mrst99}), with other uncertainties estimated to be a factor of about
$^{\times3}_{\div3}$~\cite{vkpriv}. A comparison of our measurement
with the only theoretical prediction available to date is shown in 
Fig.~\ref{fig:expvsth}. The rates of
$e^+e^-$ and $\gamma\gamma$ events with $E_T(e/\gamma) > 5$ GeV are consistent 
with those in our earlier studies~\cite{cdfee,cdfgg1}.

In conclusion, we have observed the exclusive production of two high-$E_T$
photons in proton-antiproton collisions, which constitutes the first observation
of this process in hadron-hadron collisions. The cross section is in
agreement with the only theoretical prediction, based on $g+g \rightarrow
\gamma+\gamma$, with another gluon exchanged to cancel the color and with the
$p$ and $\bar{p}$ emerging intact. If a Higgs boson exists, it should be produced 
by the same mechanism (see Fig.~\ref{fig:cepDiagrams}), and the cross sections are
related.

We thank the Fermilab staff and the technical staffs of the participating
institutions for their vital contributions. This work was supported by the U.S.
Department of Energy and National Science Foundation; the Italian Istituto
Nazionale di Fisica Nucleare; the Ministry of Education, Culture, Sports,
Science and Technology of Japan; the Natural Sciences and Engineering Research
Council of Canada; the National Science Council of the Republic of China; the
Swiss National Science Foundation; the A.P. Sloan Foundation; the
Bundesministerium f\"ur Bildung und Forschung, Germany; the Korean World Class
University Program, the National Research Foundation of Korea; the Science and
Technology Facilities Council and the Royal Society, UK; the Institut National
de Physique Nucleaire et Physique des Particules/CNRS; the Russian Foundation
for Basic Research; the Ministerio de Ciencia e Innovaci\'{o}n, and Programa
Consolider-Ingenio 2010, Spain; the Slovak R\&D Agency; the Academy of Finland;
and the Australian Research Council (ARC). We also thank V.A.~Khoze, M.G.~Ryskin, and
L.A.~Harland-Lang for many valuable discussions.


\begin{thebibliography}{99}
\bibitem{sudakov} The Sudakov factor suppresses real gluon radiation
that could fill the rapidity gaps.
\end{thebibliography}
\end{document}